\documentclass[12pt]{iopart}
\usepackage{epsf}
\usepackage{graphicx}
\usepackage{bm}
\newcommand{\simgt}{\lower.5ex\hbox{$\; \buildrel > \over \sim \;$}}
\newcommand{\simlt}{\lower.5ex\hbox{$\; \buildrel < \over \sim \;$}}

\begin{document}

\title[]{Classical and quantum radiation from a moving charge 
in an expanding universe
}


\author{Hidenori Nomura}
\address{Graduate School of Sciences, Hiroshima University, 
Higashi-hiroshima, 739-8526, Japan}

\author{ Misao Sasaki}
\address{
Yukawa Institute for Theoretical Physics, Kyoto University, 
Kyoto, 606-8502,~Japan}

\author{Kazuhiro Yamamoto}
\address{Graduate School of Sciences, Hiroshima University, 
Higashi-hiroshima, 739-8526, Japan}

\begin{abstract}
We investigate photon emission from a moving particle in an expanding
universe. This process is analogous to the radiation from an accelerated
charge in the classical electromagnetic theory.
Using the framework of quantum field theory in curved spacetime,
we demonstrate that the Wentzel-Kramers-Brillouin (WKB) approximation leads to
the Larmor formula for the rate of the radiation energy from a moving 
charge in an expanding universe. 
Using exactly solvable models in a radiation-dominated universe
and in a Milne universe, we examine the validity of the WKB 
formula. It is shown that the quantum effect suppresses 
the radiation energy in comparison with the WKB formula. 
\end{abstract}

\def\bfp{{\bm p}}
\def\bfq{{\bm q}}
\def\bfx{{\bm x}}
\def\bfk{{\bm k}}
\def\bfpf{{\bm p_{\rm f}}} 
\def\bfpi{{\bm p_{\rm i}}}
\section{Introduction}
One of the notable feature of quantum fields in curved 
spacetime is that quantum processes prohibited in the 
Minkowski spacetime is allowed \cite{BirrellD,BirrellDB}. 
For example, the emission of a photon from a moving 
massive charged particle occurs in an expanding universe, though 
such a process is prohibited by the energy momentum conservation
in the Minkowski spacetime due to the Lorentz invariance.
This subject was studied by several authors:
Pioneering work was done by Buchbinder and Tsaregorodtsev \cite{BT} 
and by Lotze\cite{Lotze}. These authors investigated the transition
probability of the process by applying QED to a
radiation-dominated universe. Then Futamase et al. and Hotta et al.
considered the case of a simple background spacetime 
with a sudden transition of the scale factor \cite{Futamase,Hotta}.
These previous works, however, focused on the transition probability 
of the process. While, in the present paper, we calculate the 
radiation energy emitted through the process. Our point of view 
is as follows: The motion of a massive charge in an expanding or
contracting universe can be regarded as an accelerated motion, because
the physical momentum of the particle decreases (increased) as the 
universe expands (contracts). Then, the photon emission process
can be regarded as the well-known classical radiation process 
from an accelerated charge \cite{Jacson}. The present 
paper aims to clarify the correspondence between the 
classical and quantum approaches to photon emission 
from a moving charge in expanding universe. 

This paper is organized as follows: In section 2, we review the
scalar QED model in the Friedmann-Robertson-Walker universe.
Then, we show that the classical radiation formula, which 
corresponds to the Larmor formula for the rate of the radiation energy 
from an accelerated charge, is derived under the WKB approximation
in the framework of quantum field theory.
In section 3, we analyze the scalar QED model without 
WKB approximation in two different universe models.
We first consider a universe
which undergoes a bounce and is radiation-dominated at
asymptotic past and future infinity. The advantage of
this model is that the equation of motion of a
 free complex scalar field is exactly solvable.
Then, the radiation energy through the photon emission process 
is numerically computed.
The result is compared with the corresponding 
result based on the WKB formula, and the condition for the validity
of the WKB formula is found.
Second, we consider the case of a bounced Milne universe,
to verify the robustness of the result. 
The last section is devoted to the summary and conclusions. 
Throughout this paper, we use the unit light velocity equals $1$.
We follow the convention $(-,+,+,+)$.

\section{Derivation of the radiation formula with the WKB approximation }
In what follows, we focus on the spatially flat Friedmann-Robertson-Walker 
spacetime, whose line element is expressed as 
\begin{eqnarray}
  ds^2=a(\eta)^2[-d\eta^2+d{\bfx}^2]=a(\eta)^2\eta_{\mu\nu}dx^\mu dx^\nu,
\label{lineelement}
\end{eqnarray}
where $\eta$ is the conformal time, and $a(\eta)$ is the scale factor.
We consider the scalar QED Lagrangian conformally 
coupled to the curvature,
\begin{eqnarray}
  {\cal S}&=& \int d^4x \sqrt{-g}
  \left[-g^{\mu\nu}\left(\nabla_{\mu}-{ieA_\mu\over \hbar }\right)
      \Phi^\dagger\left(\nabla_{\nu}+{ieA_\nu\over \hbar }\right)\Phi\right.
\nonumber
\\
  &&~~~~~~~~\left.-\left({ m^2\over \hbar^2}+\xi_{\rm conf} R\right)
  \Phi^*\Phi -{1\over 4\mu_0}F^{\mu\nu}F_{\mu\nu}\right],
\nonumber
\end{eqnarray} 
where $F_{\mu\nu}=\nabla_\mu A_\nu-\nabla_\nu A_\mu$ is the field
strength, and $\mu_0$ is the magnetic permeability of vacuum.
In this paper, we explicitly include the Planck constant $\hbar$.
Introducing the conformally rescaled field $\phi$,
\begin{eqnarray}
  \Phi={\phi\over a(\eta)},
\nonumber
\end{eqnarray}
we may rewrite the Lagrangian as 
\begin{eqnarray}
\hspace{-1cm}  {\cal S}=\int d^4x  
  \left[-\eta^{\mu\nu}\left(\partial_{\mu}-{ieA_\mu\over \hbar }\right)\phi^\dagger
                     \left(\partial_{\nu}+{ieA_\nu\over \hbar }\right)\phi
  -{m^2a(\eta)^2\over \hbar^2}\phi^* \phi 
  -{1\over 4\mu_0}f^{\mu\nu} f_{\mu\nu}\right],
\nonumber
\end{eqnarray}
where $f_{\mu\nu}=\partial_\mu A_\nu-\partial_\nu A_\mu$.
Thus the system is mathematically equivalent to the scalar 
QED in the Minkowski spacetime with the time-variable mass 
$m\,a(\eta)$. 

\begin{figure}
\begin{center}
\includegraphics[width=4.5in,angle=0]{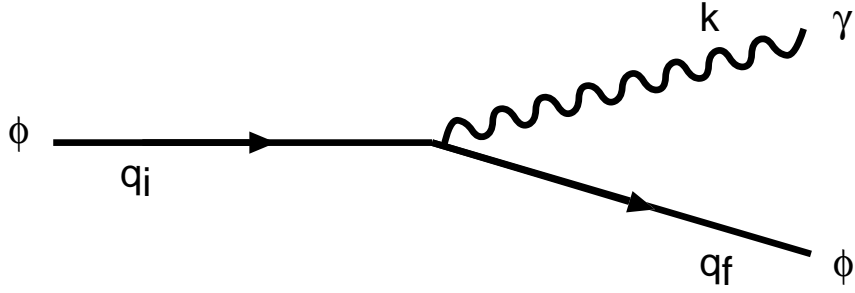}
\caption{Feynman Diagram of the photon emission process}
\label{fig1}
\end{center}
\end{figure}
We follow a general prescription for interacting fields, based
on the interaction picture approach (see e.g., \cite{BirrellD,BirrellDB}). 
We focus on the radiation energy emitted through the 
process described by the Feynman diagram in Fig.~1.
The S-matrix corresponding to the diagram is
\begin{eqnarray}
  S={-i\over \hbar}\int {{ieA^\mu\over \hbar}\left(
  \phi\partial_\mu\phi^\dagger-\phi^\dagger\partial_\mu\phi\right)}d^4x.
\label{smatrix}
\end{eqnarray}
Here the field operator in the right-hand side, $A^\mu$ and $\phi$
obey the free field equations.
The quantized free fields are expressed as follows.
For the photon field, we have 
\begin{eqnarray}
  A_\mu(x)=\sum_{r=1,2}\sum_{\bfk} \sqrt{\hbar \mu_0\over 2 \omega(k)} 
  \varepsilon_\mu^{(r)}\left\{\hat a_\bfk^{(r)} e^{-i\omega\eta}{e^{i\bfk\cdot\bfx}\over L^{3/2}}+{\rm Hermite ~Conjugate}
  \right\},
\label{photonfield}
\end{eqnarray}
where $\omega(k)=k$, $\varepsilon_\mu^{(r)}$ is the polarization vector, 
and $a_\bfk^{(r)}$ and $a_\bfk^{(r)}{}^\dagger$ are the annihilation 
and creation operators.  They satisfies the commutation relations,
\begin{eqnarray}
  &&\left[\hat a_\bfk^{(r)},\hat a_{\bfk'}^{(r')}{}^\dagger\right]
  =\delta_{r,r'}\delta_{\bfk,\bfk'}, 
\hspace{1cm}
\left[\hat a_\bfk^{(r)},\hat a_{\bfk'}^{(r')}\right]
 =\left[\hat a_\bfk^{(r)}{}^\dagger,\hat a_{\bfk'}^{(r')}{}^\dagger\right]=0.
\nonumber
\end{eqnarray}
The vacuum state of the photon field is
\begin{eqnarray}
  \hat a_\bfk^{(r)}|0\big>^{(1)}=0, ~\hspace{0.5cm}\mbox{for any}~\bfk,~r.
\end{eqnarray}

The complex scalar field is quantized as 
\begin{eqnarray}
  \phi(x)=\sqrt{\hbar}\sum_{\bfq} 
    \left\{\hat b_\bfq \varphi_\bfq(\eta) {e^{i\bfq\cdot\bfx}\over L^{3/2}}
        +\hat c_\bfq^\dagger \varphi_\bfq^*(\eta) {e^{-i\bfq\cdot\bfx}\over L^{3/2}}
  \right\},
\label{complexscalar}
\end{eqnarray}
where $\hat b_\bfq$ and $\hat b_\bfq^\dagger$ satisfy
\begin{eqnarray}
  &&\left[\hat b_{\bfq},\hat b_{\bfq'}^\dagger\right]
  =\delta_{\bfq,\bfq'},
\hspace{1cm}
  \left[\hat b_\bfq,\hat b_{\bfq'}\right]
   =\left[\hat b_\bfq^\dagger,\hat b_{\bfq'}^\dagger\right]=0, 
\label{commutationrelation}
\end{eqnarray}
and $\hat c_\bfq$ and $\hat c_\bfq^\dagger$ satisfy the same
commutation relations. The mode function $\varphi_\bfq(\eta)$
satisfies the equation of motion 
\begin{eqnarray}
  \left({d^2\over d\eta^2}+{m^2a(\eta)^2\over \hbar^2}
      +\bfq^2\right)\varphi_\bfq(\eta)=0,
\label{equationofmotion}
\end{eqnarray}
with the normalization condition
\begin{eqnarray}
  {d\varphi_\bfq^*(\eta)\over d\eta}\varphi_\bfq(\eta)-
   \varphi_\bfq^*(\eta){d\varphi_\bfq(\eta)\over d\eta}=i.
\end{eqnarray}
The vacuum state of the complex scalar field is
\begin{eqnarray}
  \hat b_\bfq|0\big>^{(2)}=\hat c_\bfq|0\big>^{(2)}=0,
~\hspace{0.5cm}{\rm for~any}~\bfq.
\end{eqnarray}
It may be noted that because of the time-variation of
the mass $m\,a(\eta)$ the definition of the vacuum state
can be ambiguous. In this section, however, we assume that
there exists a natural stable vacuum state, which is the case
when the WKB approximation is valid.

With the WKB approximation, the mode function $\varphi_\bfq(\eta)$ is 
given as
\begin{eqnarray}
  \varphi_\bfq(\eta)=\sqrt{1\over 2\Omega_\bfq(\eta)}\exp
  \left[-i\int_{\eta_*}^\eta {\Omega_\bfq(\eta')} d\eta'
  \right],
\end{eqnarray}
where
\begin{eqnarray}
  \Omega_\bfq(\eta)={\sqrt{m^2a(\eta)^2+\hbar^2\bfq^2}\over\hbar}.
\end{eqnarray}
We can write the condition that the WKB formula is valid, as (see e.g., \cite{BirrellD})
\begin{eqnarray}
  \Omega_{\bfq}^2\gg {1\over 2}\Biggl|{\ddot \Omega_\bfq\over \Omega_\bfq}
 -{3\over 2}{\dot \Omega_\bfq^2\over \Omega_\bfq^2}\Biggr|,
\label{wkbcondition}
\end{eqnarray}
where the dot means the differentiation with respect to $\eta$.

Using the WKB mode functions, 
we evaluate the transition amplitude of the process described by Fig.~1, 
\begin{eqnarray}
{\rm Transition~ Amplitude}=\left<{\rm f}| S |{\rm i}\right>,
\label{transitionamplitude}
\end{eqnarray}
where the initial state and the final state are 
\begin{eqnarray}
  &&| {\rm i}\big>=\hat b_{\bfq_i}^\dagger |0\big>^{(1)}|0\big>^{(2)},
 \\
  &&| {\rm f}\big>=\hat b_{\bfq_f}^\dagger \hat a_{\bfk}^{(r)}{}^\dagger
    |0\big>^{(1)}|0\big>^{(2)},
\end{eqnarray}
respectively.

The radiation energy is 
\begin{eqnarray}
  E={(2\pi)^3\over L^3} \sum_{r=1,2} \sum_\bfk \sum_\bfq \hbar \omega(k)
  |\rm Transition ~Amplitude|^2,
\end{eqnarray}
which gives
\begin{eqnarray}
  E={2{e^2}\over \varepsilon_0} \int {d^3\bfk \over (2\pi)^3} \left(
  \bfq_i^2-{(\bfq_i\cdot \bfk)^2\over k^2}\right) \left|
  \int d\eta e^{ik\eta} \varphi_{\bfq_f}^*(\eta)\varphi_{\bfq_i}(\eta)
  \right|^2,
\label{larmora}
\end{eqnarray}
where  $\bfq_f=\bfq_i-\bfk$, and $ \varepsilon_0$ is the
permittivity of vacuum, which is related to $\mu_0$ 
as $ \varepsilon_0 \mu_0=1/c^2=1$.
We set $\bfq_i\cdot \bfk=q_ik\cos\theta$, and rewrite 
(\ref{larmora}) as
\begin{eqnarray}
\hspace{-1cm}  &&E={e^2\over 2\varepsilon_0} 
   \int {d^3\bfk \over (2\pi)^3} \bfq_i^2\left(1-\cos^2\theta\right) 
\nonumber\\
  &&\times\left|  \int d\eta
  {1\over \sqrt{\Omega_{\bfq_i}(\eta)\Omega_{\bfq_f}(\eta)}}
   \exp\left[ik\eta+i\int_{\eta_*}^\eta d\eta'\Omega_{\bfq_f}(\eta')
            -i\int_{\eta_*}^\eta d\eta'\Omega_{\bfq_i}(\eta')
  \right]
  \right|^2,
\nonumber\\
\end{eqnarray}
where
\begin{eqnarray}
  p^0=\hbar\Omega_\bfq(\eta)
  =\sqrt{\hbar^2\bfq^2+m^2a(\eta)^2}
  =\sqrt{\bfp^2+m^2a(\eta)^2},
\end{eqnarray}
where the first and last equalities
follow from the definition of the momentum of the charged particle,
\begin{eqnarray}
  \bfp=\hbar\bfq.
\end{eqnarray}

We follow the prescription in ref. \cite{Higuchi},
\begin{eqnarray}
  \int_{\eta_*}^\eta d\eta'\left(
  \Omega_{\bfp_f}(\eta')-\Omega_{\bfq_i}(\eta')
  \right)
  &\simeq&
  \int_{\eta_*}^\eta d\eta'\left(
  \bfq_f-\bfq_i
  \right)\cdot {\partial \over \partial\bfq}\sqrt{\bfq^2+{m^2a(t)^2\over \hbar^2}}
  \Bigg|_{\bfp=\bfp_i}\\
\nonumber
  &=&-\int_{\eta_*}^\eta d\eta' {\bfk\cdot \bfp_i \over p^0},
\end{eqnarray}
and 
\begin{eqnarray}
  \sqrt{\Omega_{\bfq_i}(\eta)\Omega_{\bfq_f}(\eta)}\simeq\Omega_{\bfq_i}(\eta)={p^0\over \hbar},
\end{eqnarray}
where we used 
$
\bfq_f=\bfq_i-\bfk
$
and 
\begin{eqnarray}
k(=|\bfk|)\ll q_f(=|\bfq_f|),~~{\rm and}~q_i(=|\bfq_i|).  
\label{classicalpic}
\end{eqnarray}
Then, we have
\begin{eqnarray}
  E&=&{e^2\over 2\varepsilon_0} 
   \int_0^\infty {dk k^2 \over (2\pi)^2}\int_{-1}^1d\cos\theta  \left(1-\cos^2\theta\right) 
\nonumber\\
  &&\times
  \left|  \int d\eta
  {p_i\over p^0(\eta)}
   \exp\left[ik\int_{\eta_*}^\eta d\eta'\left( 1-{p_i\cos\theta\over p^0(\eta')}\right)
  \right]
  \right|^2.
\end{eqnarray}

We can choose ${\bfp}_i$ to be in proportion to $z$-axis. 
Therefore, we may write
\begin{eqnarray}
  p_i=p^z={dz\over d\tau}, \qquad p^0={d\eta\over d\tau},
\end{eqnarray}
where $\tau$ is the parameter chosen as
\begin{eqnarray}
  \eta_{\mu\nu}{dx^{\mu}\over d\tau}{dx^{\nu}\over d\tau}=-(p^0)^2+p_z^2
    =-m^2a^2(\eta).
\label{affine}
\end{eqnarray}
Because we may write $p_i/p^0=dz/d\eta$, we have
\begin{eqnarray}
  &&\left|  \int d\eta
  {p_i\over p^0(\eta)}
   \exp\left[i\int_{\eta_*}^\eta
   d\eta'k\left( 1-{p_i\cos\theta\over p^0(\eta')}\right)
  \right]
  \right|^2
\nonumber\\
  && ~~~~~~~~~~~~~~~~~~
 =\left|  \int dz \exp\left[ ik(\eta(z)-\cos\theta z)\right]  \right|^2.
\label{affine2}
 \end{eqnarray}

Instead of $z$ or $\eta$, introducing the variable $\xi$ defined by
\begin{eqnarray}
  \xi\equiv\eta-\cos\theta z(\eta),
\end{eqnarray}
we have
\begin{eqnarray}
  E={e^2\over 2\varepsilon_0} 
   \int_0^\infty {dk k^2 \over (2\pi)^2}\int_{-1}^1d\cos\theta  \left(1-\cos^2\theta\right) 
  \left|  \int d\xi{dz\over d\xi} e^{ik\xi } \right|^2.
\end{eqnarray}
Furthermore, the partial integration gives
\begin{eqnarray}
  \Biggl|\int d\xi{dz\over d\xi}e^{ik\xi}\Biggr|^2
    ={1\over k^2}\Biggl|\int d\xi{d^2z\over d\xi^2}e^{ik\xi}\Biggr|^2,
\end{eqnarray}
where we have assumed that the boundary term has no contribution.
The neglect of the boundary term is justified
if $|dz/d\xi|$ is finite at all times. Since we have
\begin{eqnarray}
\frac{dz}{d\xi}
=\frac{\dot z}{1-\cos\theta \dot z}\,;
\quad
|\dot z|=\frac{|p_z|}{\sqrt{p_z^2+m^2a^2}}<1\,,
\end{eqnarray}
where $\dot z=dz/d\eta$,
we see that $|dz/d\xi|$ is bounded at all times.

The integration with respect to $k$ yields
\begin{eqnarray}
  E={e^2\over 8\pi\varepsilon_0}\int_{-1}^{1}d\cos\theta(1-\cos^2\theta)
    \int d\xi\Biggl({d^2z\over d\xi^2}\Biggr)^2,
\label{energyy}
\end{eqnarray}
where the integration is extended to $-\infty\leq k\leq\infty$,
and divided by the factor 2 \cite{Higuchi}.
From the definition of $\xi$, we have
\begin{eqnarray}
  {d\xi \over d\eta}=(1-\dot{z}\cos\theta),~~~~~~
  {d^2z\over d\xi^2}={\ddot{z}\over (1-\dot{z}\cos\theta)^3}\,.
\end{eqnarray}
Then, Eq.~(\ref{energyy}) is rephrased as
\begin{eqnarray}
  E={e^2\over 8\pi\varepsilon_0}\int_{-1}^{1}d\cos\theta(1-\cos^2\theta)
    \int d\eta{\ddot{z}^2\over (1-\dot{z}\cos\theta)^5}.
\end{eqnarray}
Finally, the integration with respect to $\cos\theta$ yields
\begin{eqnarray}
  E={e^2\over 6\pi\varepsilon_0}\int d\eta{\ddot{z}^2\over (1-\dot{z}^2)^3}.
\label{fene}
\end{eqnarray}
This result is the same as the Larmor formula in the case when 
the particle moves along a straight line \cite{Jacson}. 
In the present case, Eq.~(\ref{fene}) is rewritten as
\begin{eqnarray}
    E={e^2p_{i}^2\over 6\pi \varepsilon_0m^2}\int d\eta{\dot{a}^2\over a^4}.
\label{feneb}
\end{eqnarray}
Note that this is the energy in the conformally rescaled spacetime,
which is not the physical energy. From this  
expression, however, we can read 
the physical radiation energy as ${\cal E}=E/a$, and 
the physical rate of the 
radiation energy per unit time as 
\begin{eqnarray}
  {d{\cal{E}}\over dt}={1\over a^2}{dE\over d\eta}
    ={e^2p_{phys}^2H^2\over 6\pi\varepsilon_0 m^2},
\end{eqnarray}
where $t$ is the cosmic time, $p_{phys}(=p_i/a)$ is the physical
momentum and $H$ is the Hubble expansion rate. This result is 
consistent with that in ref.~\cite{Futamase}, which is 
obtained from consideration of the classical electromagnetic
radiation formula of a moving charge in an expanding universe. 

Finally in this section, let us summarize the necessary
condition for reproducing the radiation formula in the 
classical electromagnetic theory.  We started with the
WKB formula of the mode function, for which 
Eq.~(\ref{wkbcondition}) is needed. In addition, we assumed 
Eq.~(\ref{classicalpic}). Although this condition is independent of the 
necessary condition for the WKB approximation, Eq.~(\ref{wkbcondition}),
we assumed it because this additional assumption is necessary
to recover the conventional picture for the classical radiation 
from a charged massive particle, in which the massive field should behave
like a particle and the photon field should behave like a wave.


\begin{figure}
\begin{center}
\includegraphics[width=3.5in,angle=0]{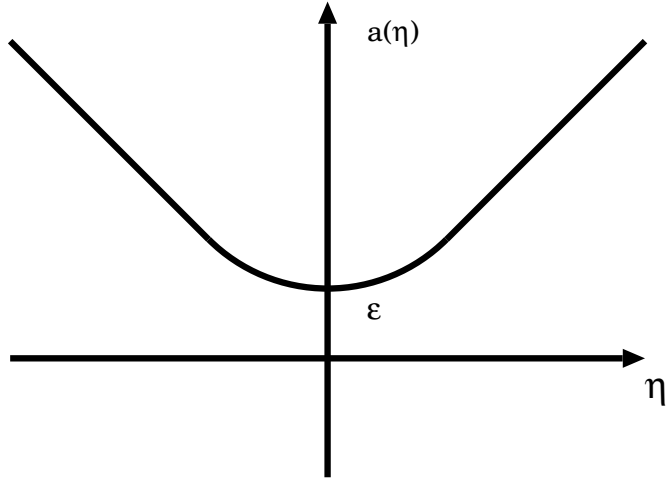}
\caption{The scale factor of the bounced radiation dominant universe
as a function $\eta$.}
\label{fig2}
\end{center}
\end{figure}

\def\alphaa{{\varrho}}
\section{Quantum radiation formula}
In this section, we calculate the radiated energy without 
the WKB approximation. We start with reviewing the 
formalism. Now we return to the S-matrix expression~(\ref{smatrix}).
A characteristic feature of the quantum field theory in 
curved spacetime is that vacuum state may not be stable.
Namely, the vacuum state may change, and the particle
creation phenomenon can occur. This vacuum effect is 
not taken into account in the analysis based on the WKB 
approximation in the previous section. 
In this section, we adopt the formalism for evaluating 
the S-matrix element, taking the vacuum effect into account.

For definiteness, we assume that the interaction is switched off 
in the asymptotic past infinity ($\eta=-\infty$) and future infinity
($\eta=+\infty$), 
where the different vacuum states $|0\big>_{\rm in}^{(2)}$ and 
$|0\big>_{\rm out}^{(2)}$, respectively, are defined for the free 
field. Then, similar to Eq.~(\ref{complexscalar}), we may 
write the quantized field as 
\begin{eqnarray}
  \phi(x)=\sqrt{\hbar}\sum_{\bfq} 
    \left\{\hat b^{\rm in}_\bfq \varphi^{\rm in}_\bfq(\eta) 
 {e^{i\bfq\cdot\bfx}\over L^{3/2}}
        +\hat c^{\rm in}_\bfq{}^\dagger \varphi^{\rm in}_\bfq{}^*(\eta) 
 {e^{-i\bfq\cdot\bfx}\over L^{3/2}}
  \right\},
\label{phiin}
\end{eqnarray}
or as
\begin{eqnarray}
  \phi(x)=\sqrt{\hbar}\sum_{\bfq} 
    \left\{\hat b^{\rm out}_\bfq \varphi^{\rm out}_\bfq(\eta) 
 {e^{i\bfq\cdot\bfx}\over L^{3/2}}
        +\hat c^{\rm out}_\bfq{}^\dagger \varphi^{\rm out}_\bfq{}^*(\eta) 
 {e^{-i\bfq\cdot\bfx}\over L^{3/2}}
  \right\},
\label{phiout}
\end{eqnarray}
where $b^{\rm in}_\bfq$, $c^{\rm in}_\bfq$ and 
$b^{\rm in}{}^\dagger_\bfq$, $c^{\rm in}{}^\dagger_\bfq$ are the
annihilation and the creation operators, respectively, 
with respect to the in-vacuum at $\eta=-\infty$, 
while $b^{\rm out}_\bfq$, $c^{\rm out}_\bfq$ and 
$b^{\rm out}{}^\dagger_\bfq$, $c^{\rm out}{}^\dagger_\bfq$
are those with respect to the out-vacuum at $\eta=+\infty$.
The in-vacuum and out-vacuum states are expressed as
\begin{eqnarray}
 &&{\hat b^{\rm in}_\bfq} |0\big>_{\rm in}^{(2)}
={\hat c^{\rm in}_\bfq} |0\big>_{\rm in}^{(2)}=0,
\end{eqnarray}
and
\begin{eqnarray}
 &&{\hat b^{\rm out}_\bfq}|0\big>_{\rm out}^{(2)}
={\hat c^{\rm out}_\bfq}|0\big>_{\rm out}^{(2)}=0,
\end{eqnarray}
for any $\bfq$.
These annihilation and the creation operators satisfy the 
same commutation relations as Eq.~(\ref{commutationrelation}).

In Eqs. (\ref{phiin}) and (\ref{phiout}), $\varphi_\bfq^{\rm in}(\eta)$
and $\varphi_\bfq^{\rm out}(\eta)$ are the mode functions with respect
to the vacuum states $|0\big>_{\rm in}^{(2)}$ and $|0\big>_{\rm out}^{(2)}$, 
respectively. They are related by the Bogoliubov transformation,
\begin{eqnarray}
  &&\varphi_\bfq^{\rm in}(\eta)=\alpha_\bfq \varphi_\bfq^{\rm out}(\eta)
  +\beta_\bfq \varphi_\bfq^{\rm out}{}^*(\eta)
\end{eqnarray}
where $\alpha_\bfq$ and $\beta_\bfq$ satisfy
the normalization condition
\begin{eqnarray}
   |\alpha_\bfq|^2-|\beta_\bfq|^2=1.
\end{eqnarray}
The creation and annihilation operators are related as
\begin{eqnarray}
  &&\hat b^{\rm in}_\bfq=\alpha_\bfq^* \hat b^{\rm out}_\bfq
                   -\beta_\bfq^* \hat c^{\rm out}_{-\bfq}{}^\dagger, 
\\
  &&\hat c^{\rm in}_{-\bfq}{}^\dagger=\alpha_\bfq \hat c^{\rm out}_{-\bfq}{}^\dagger
                   -\beta_\bfq \hat b^{\rm out}_{\bfq}.
\end{eqnarray}
Using the above relations, we have
\begin{eqnarray}
  &&\varphi_\bfq^{\rm out}{}^*(\eta) \hat b^{\rm out}_\bfq{}^\dagger
   +\varphi_\bfq^{\rm out}{}  (\eta) \hat c^{\rm out}_{-\bfq}
  =
  {\varphi_\bfq^{\rm in}{}^* (\eta)\over \alpha_\bfq^* } 
                             \hat b^{\rm out}_\bfq{}^\dagger
 +{\varphi_\bfq^{\rm out}    (\eta)\over \alpha_\bfq^*}
                             \hat c^{\rm in}_{-\bfq}
\end{eqnarray}
and
\begin{eqnarray}
  &&\varphi_\bfq^{\rm in}{}     (\eta) \hat b^{\rm in}_\bfq{}
   +\varphi_\bfq^{\rm in}{}^*  (\eta)\hat c^{\rm in}_{-\bfq}{}^\dagger
  =
  {\varphi_\bfq^{\rm out}{}  (\eta) \over \alpha_\bfq^* } 
                             \hat b^{\rm in}_\bfq{}
 +{\varphi_\bfq^{\rm in}{}^*    (\eta)\over \alpha_\bfq^*}
                             \hat c^{\rm out}_{-\bfq}{}^\dagger.
\end{eqnarray}
As for the photon field, because of its conformal invariance,
the vacuum state is invariant in time, 
$|0\big>^{(1)}=|0\big>^{(1)}_{\rm in}=|0\big>^{(1)}_{\rm out}$.

The transition amplitude in the lowest order of $e$ 
is evaluated in the similar form as 
Eq.~(\ref{transitionamplitude}), but with 
\begin{eqnarray}
  &&| {\rm i}\big>_{\rm in}=\hat b_{\bfq_i}^{\rm in}{}^\dagger |0\big>^{(1)}
  |0\big>^{(2)}_{\rm in},
 \\
  &&| {\rm f}\big>_{\rm out}= \hat b_{\bfq_f}^{\rm out}{}^\dagger \hat a_{\bfk}^{(r)}{}^\dagger
    |0\big>^{(1)}  |0\big>^{(2)}_{\rm out}.
\end{eqnarray}
In the computation of the transition amplitude, it is necessary
to regularize the divergence arising from the vacuum-to-vacuum
amplitude. In the flat background, this can be done unambiguously
by taking the normal-order product of operators. However, in a curved
spacetime, there arises ambiguity because the in-state annihilation/creation
 operators are different from the out-state annihilation/creation operators.
In fact, there will be particle creation from vacuum and the
vacuum-to-vacuum amplitude will not longer be unity any more.

To deal with this situation properly, we consider
the {\it generalized normal product} of operators, which is defined
as the form where the operators are expressed only in terms of the 
in-state annihilation operator and the out-state creation operators, 
and all the out-state creation operator are placed to the 
left of all the in-state annihilation operators \cite{BFG}.
This is adopted in ref.~\cite{BT}.
The nice properties of this generalized normal ordering are that
it is symmetrically defined with respect to 
the in-state and out-state operators, and 
that the vacuum-to-vacuum amplitude is normalized to unity.
In particular, this latter property means that it
minimizes the effect of particle creation from vacuum. 
Since what we are interested in is not the vacuum particle creation
but the transition amplitude for a massive particle
to radiate under the electromagnetic interaction, we adopt the
 generalized normal ordering to regularize the divergence. 
We note that what would be actually observed
will be inevitably contaminated by the effect of particle creation.
However, whether there is a way to separate out this vacuum effect 
observationally is beyond the scope of the
present paper.

Following ref.\cite{BT}, we define 
\begin{eqnarray}
  {\rm Transition~amplitude}= {\displaystyle{
  {}_{\rm out}\bigl<{\rm f}|N[S]|{\rm i}\big>_{\rm in}}
  \over \displaystyle{{}_{\rm out}\bigl<0|0\big>_{\rm in}}}.
\end{eqnarray}
Here $|0\big>_{\rm in}=|0\big>^{(1)}|0\big>^{(2)}_{\rm in}$ and
     $|0\big>_{\rm out}=|0\big>^{(1)}|0\big>^{(2)}_{\rm out}$,
respectively, $N[S]$ denotes the generalized normal 
product of the S-matrix, and ${}_{\rm out}\bigl<0|0\bigr>_{\rm in}$ is 
the vacuum to vacuum amplitude.

We then find the formula for the radiation energy as 
\begin{eqnarray}
  E={2e^2{{\bfq}_i}^2
   \over (2\pi)^2\varepsilon_0}\int_{0}^{\infty} dk k^2
    \int_{-1}^{1}d\cos\theta
    {1-\cos^2\theta\over |\alpha_{{\bfq}_i}|^2|\alpha_{{\bfq}_f}|^2}
    \Biggl|\int d\eta e^{ik\eta}
    \varphi^{in*}_{{\bfq}_f}\varphi^{\rm out}_{{\bfq}_i}\Biggr|^2.
\label{gnoene}
\end{eqnarray}

\begin{figure}
\begin{center}
\includegraphics[width=3.3in,angle=0]{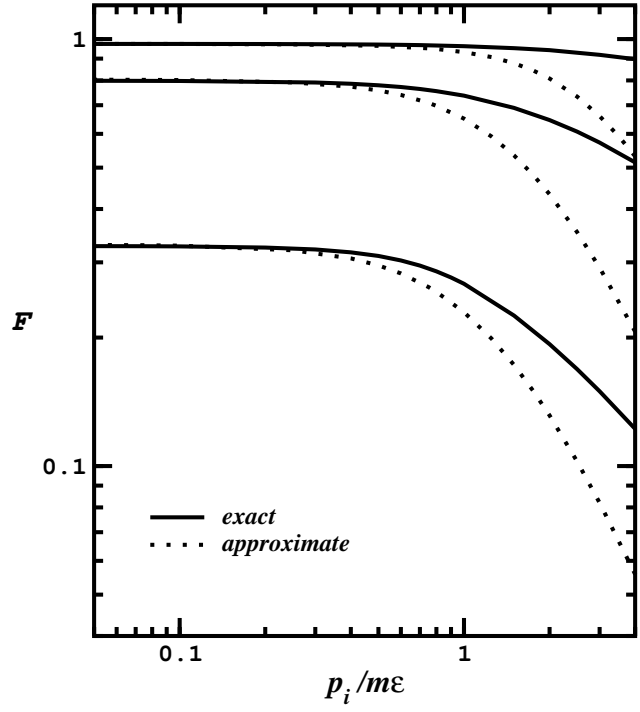}
\caption{$F$ as function of $p_i/m\epsilon$ with 
fixed as $\hbar\alphaa/m\epsilon^2=0.01, ~0.1, ~1$,
from top to bottom. The solid curve is exact, 
Eq.~(\ref{functionfexact}), while the dotted 
curve is approximate, Eq.~(\ref{besselk}).
}
\label{fig3}
\end{center}
\end{figure}

\begin{figure}
\begin{center}
\includegraphics[width=3.5in,angle=0]{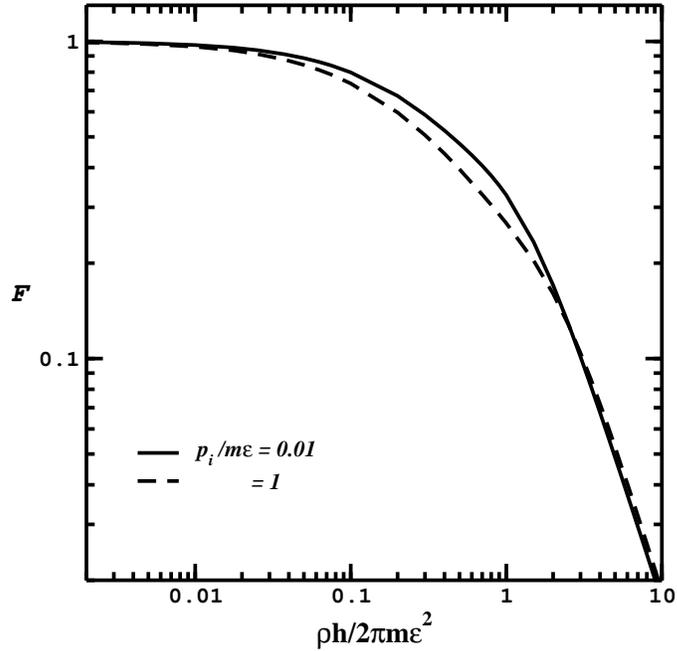}
\caption{$F$ as function of $\hbar \alphaa/m\epsilon^2$ 
with fixed as $p_i/m\epsilon=0.01$(dashed curve) and $1$(solid curve). 
}
\label{fig4}
\end{center}
\end{figure}

\subsection{Example I}

In this subsection, we consider a time-symmetric bounce universe
which asymptotically approaches a contracting and expanding 
radiation-dominated universe (see Fig.~2). The scale factor is
given in terms of the conformal time $\eta$ by \cite{AS,ASb}
\begin{eqnarray}
  a(\eta)=\sqrt{\alphaa^2\eta^2+\epsilon^2} \qquad 
    (-\infty<\eta<\infty),
\label{scalefactor}
\end{eqnarray}
which recovers a radiation-dominated Friedmann universe
in the asymptotic regions,
$a(t)\propto t^{1/2} ~(\eta\rightarrow \pm\infty)$.
In this background spacetime, the WKB formula 
(\ref{fene}) gives
\begin{eqnarray}
  E_{cl}={e^2p_i^2\over 6\pi \varepsilon_0m^2}\int\nolimits_{-\infty}^{\infty}
    d\eta{\alphaa^4\eta^2\over (\alphaa^2\eta^2+\epsilon^2)^3}
    ={e^2p_i^2\alphaa\over 48\varepsilon_0m^2\epsilon^3}.
\end{eqnarray}

Equation of motion (\ref{equationofmotion}) is reduced to 
the Weber's differential equation
\begin{eqnarray}
  {d^2\varphi\over dz^2}+\Bigl(\nu+{1\over 2}
    -{1\over 4}z^2\Bigr)\varphi=0,
\end{eqnarray}
where
\begin{eqnarray}
  \nu={1\over 2}\Bigl({{\bfq}^2+m^2\epsilon^2/\hbar^2\over \pm im\alphaa/\hbar}-1\Bigr),
    \qquad z=\pm(1\pm i)\sqrt{m\alphaa/\hbar}\eta.
\end{eqnarray}
Therefore, the mode functions are constructed as \cite{Kluger}
\begin{eqnarray}
  \varphi^{\rm in}_{{\bfq}}(\eta)
    ={1\over (2m\alphaa/\hbar)^{1/4}}e^{-\pi\lambda/4}
    D_{i\lambda-1/2}(-(1-i)\sqrt{m\alphaa/\hbar}\eta),
\label{cylindera}
\\
  \varphi^{\rm out}_{{\bfq}}(\eta)
    ={1\over (2m\alphaa/\hbar)^{1/4}}e^{-\pi\lambda/4}
    D_{-i\lambda-1/2}((1+i)\sqrt{m\alphaa/\hbar}\eta),
\label{cylinderb}
\end{eqnarray}
with
\begin{eqnarray}
  \lambda={{\bfq}^2+m^2\epsilon^2/\hbar^2\over 2m\alphaa/\hbar}.
\end{eqnarray}
Using the mathematical formula for the parabolic cylinder
function,
\begin{eqnarray}
  D_{\nu}(z)={\sqrt{2\pi}\over \Gamma(-\nu)}e^{\pm i\pi(\nu+1)/2}
    D_{-\nu-1}(\mp iz)+e^{\pm i\pi\nu}D_{\nu}(-z).
\end{eqnarray}
we easily find the Bogoliubov coefficients
\begin{eqnarray}
  \alpha_{\bfq}={\sqrt{2\pi}\over \Gamma(-i\lambda+1/2)}
    e^{\pi(-\lambda+i/2)/2}, \qquad
    \beta_{\bfq}=e^{\pi(-\lambda-i/2)}.
\end{eqnarray}

Using the mathematical formula for the parabolic cylinder function
\cite{Nikishov},
\begin{eqnarray}
  &&\hspace{-10mm}\int\nolimits_{-\infty}^{\infty}d\xi e^{-i\rho\xi}
    D_{-i\beta}((1+i)\xi)D_{-i\alpha}(-(1+i)\xi) \nonumber \\
   &&=\sqrt{\pi}\exp\Bigl[-{\pi\over 4}i
    +{3\pi\over 4}(\alpha-\beta)
    +{i\over 4}\rho^2+i(\alpha-\beta)\ln{\rho\over \sqrt{2}}\Bigr]
\nonumber\\
  &&\times    U\Bigl(i\alpha, 1+i(\alpha-\beta),
     -i{\rho^2\over 2}\Bigr),
\end{eqnarray}
we have
\begin{eqnarray}
    &&\int\nolimits_{-\infty}^{\infty}d\eta e^{ik\eta}
    \varphi_{{\bfp}_i}^{\rm out}(\eta)\varphi_{{\bfp}_f}^{\rm in*}(\eta)
\nonumber\\
  &&    ={\sqrt{\pi}\hbar\over \sqrt{2}m\alphaa}
    e^{-\pi(\lambda+\bar\lambda)/4}
    \exp\Bigl[-{\pi\over 4}i+{3\pi\over 4}(\lambda-\bar\lambda)
    +{i\over 4}{k^2\over m\alphaa/\hbar}
    +i(\lambda-\bar\lambda)\ln{k\over \sqrt{2m\alphaa/\hbar}}\Bigr] 
    \nonumber \\
   &&\hspace{40mm}\times U\Bigl(i\lambda+{1\over 2}, 1+(\lambda-\bar\lambda),
    -i{k^2\over 2m\alphaa/\hbar}\Bigr),
\end{eqnarray}
where $U(a, c, z)$ is the second order confluent hypergeometric
function \cite{Magnus}, and we used the notation
\begin{eqnarray}
  \lambda&=&{{\bfq_i}^2+m^2\epsilon^2/\hbar^2\over 2m\alphaa/\hbar}
  ={1\over 2}{m \epsilon^2\over \alphaa\hbar}\left(1+{{\bfp}_i^2\over m^2\epsilon^2}\right),
\\
  \bar\lambda&=&{{(\bfq_i-k)}^2+m^2\epsilon^2/\hbar^2\over 2m\alphaa/\hbar}
  ={1\over 2}{m \epsilon^2\over \alphaa\hbar}\left(1+{({\bfp}_i-\hbar{\bfk})^2\over m^2\epsilon^2}\right)
\nonumber\\
  &=&{1\over 2}{m \epsilon^2\over \alphaa\hbar}\left(1+
  {{\bfp}_i^2\over m^2\epsilon^2}
  -2{{p}_i\over m\epsilon}{\hbar \alphaa\over m\epsilon^2} \hat k \cos\theta
  +{\hbar^2 \alphaa^2\over m^2\epsilon^4}\hat k^2
\right).\nonumber
\\
\end{eqnarray}

Then, the radiation energy $E$ is represented by
\begin{eqnarray}
  E=E_{cl}\times F\Bigl({p_i\over \epsilon m},{\hbar\alphaa\over \epsilon^2m}\Bigr),
\end{eqnarray}
where
\begin{eqnarray}
  &&\hspace{-20mm}F\Bigl({p_i\over \epsilon m},
    {\hbar\alphaa\over \epsilon^2m}\Bigr)
    ={12\over \pi}
    \int\nolimits_0^{\infty}d\hat{k}\hat{k}^2
    \int\nolimits_{-1}^1d\cos\theta
    {1-\cos^2\theta\over (1-e^{-2\pi\lambda})(1-e^{-2\pi\bar\lambda})}
    e^{\pi(\lambda-2\bar\lambda)}
    \nonumber \\ 
   &&\hspace{50mm}\times\Bigl|U\Bigl(i\lambda+{1\over 2}, 
    1+i(\lambda-\bar\lambda)
    ,-i{\hbar \alphaa\hat{k}^2\over 2m\epsilon^2}\Bigr)\Bigr|^2,
\label{functionfexact}
\end{eqnarray}
and we have defined $\hat k=(\epsilon/\alphaa){k}$.
Note that the function $F$ describes
the deviation from the WKB formula.

Now let us consider the classical limit by taking the limit 
$\hbar \rightarrow 0$. We use the mathematical formula
\begin{eqnarray}
  \lim_{a\to\infty}U(a,c,z/a)
    ={2z^{(1-c)/2}K_{c-1}(2\sqrt{z})\over \Gamma(a+1-c)},
\end{eqnarray}
where $K_{\nu}(z)$ is the modified Bessel function. Then, we have
\begin{eqnarray}
  F\Bigl({p_{i}\over m\epsilon},
    {\hbar\alphaa\over m\epsilon^2}\Bigr)
    &\simeq&{24\over \pi^2}\int\nolimits_{0}^{\infty}d\hat{k}\hat{k}^2
    \int\nolimits_{-1}^1 dx(1-x^2)e^{\pi(\lambda-\bar{\lambda})}
\nonumber\\
  &&\times    \Biggl|K_{i(\lambda-\bar{\lambda})}
    \Biggl(\hat{k}\sqrt{1+{p_{i}^2\over m^2\epsilon^2}}\Biggr)\Biggr|^2
\label{besselk}
\end{eqnarray}
with 
\begin{eqnarray}
  \lambda-\bar\lambda={p_i\over m\epsilon}
\hat k x- {1\over 2}{\hbar \alphaa\over m\epsilon^2}\hat k^2.
\end{eqnarray}
We have evaluated the function $F$ numerically. The cases of
$p_i/m\epsilon=0.01$ and $1$ are shown in Fig.~\ref{fig4}.
In all cases we analyzed, we found $F$ is a decreasing function
of $\hbar\varrho/m\epsilon^2$.

In the limit $\hbar \rightarrow 0$, 
\begin{eqnarray}
  \lambda-\bar\lambda={p_i\over m\epsilon}\hat k x=s\,.
\end{eqnarray}
Then, we have
\begin{eqnarray}
\lim_{\hbar\rightarrow0} F
   ={24\over \pi^2}\int\nolimits_{0}^{\infty}d\hat{k}\hat{k}^2
    \int\nolimits_{-1}^1 dx(1-x^2)e^{\pi s}
    \Biggl|K_{is}
    \Biggl(\hat{k}\sqrt{1+{p_{i}^2\over m^2\epsilon^2}}\Biggr)\Biggr|^2.
\label{besselkk}
\end{eqnarray}
Thus $F$ is the function of only $p_i/m\epsilon$ in this limit.
{}By numerical analysis of the right-hand side of Eq.~(\ref{besselkk}), 
we find
\begin{eqnarray}
\lim_{\hbar\rightarrow0}F=1,
\label{besselkkd}
\end{eqnarray}
irrespectively of $p_i/m\epsilon$. We therefore infer that the integral 
of Eq.~(\ref{besselkk}) gives $1$.
 Although we have not yet succeeded to show it in general,
we can show that, for the case $p_i/m\epsilon\ll1$, Eq.~(\ref{besselkk}) yields
\begin{eqnarray}
  \lim_{\hbar\rightarrow0}F\simeq{24\over \pi^2}\int\nolimits_{-1}^1 dx(1-x^2)
    \int\nolimits_0^{\infty}d\hat{k}
\hat{k}^2K_0(\hat{k})^2=1,
\end{eqnarray}
by using the integral formula for the modified Bessel function \cite{Magnus},
\begin{eqnarray}
  &&\hspace{-10mm}\int\nolimits_0^{\infty}x^{-\lambda}K_{\mu}(ax)
    K_{\nu}(bx)dx \nonumber \\
   &&={2^{-2-\lambda}a^{-nu+\lambda-1}b^{\nu}\over \Gamma(1-\lambda)}
    \Gamma\Bigl({1-\lambda+\mu+\nu\over 2}\Bigr)
    \Gamma\Bigl({1-\lambda-\mu+\nu\over 2}\Bigr)
\nonumber\\
  &&\hspace{30mm}\times    \Gamma\Bigl({1-\lambda+\mu-\nu\over 2}\Bigr)
    \Gamma\Bigl({1-\lambda-\mu-\nu\over 2}\Bigr) \nonumber \\
   &&\hspace{30mm}\times F\Bigl(
    {1-\lambda+\mu+\nu\over 2},{1-\lambda-\mu+\nu\over 2}
    ;1-\lambda;1-{b^2\over a^2}\Bigr).
\end{eqnarray}
This demonstrates that the exact formula in the limit of ${\hbar\rightarrow0}$ 
agrees with the WKB approximate formula. Then, the decrease of $F$ from $1$
comes from the term in proportion to $\hbar$, hence the suppression is 
understood as the quantum effect. 

Let us summarize the result. 
The quantum radiation formula agrees with the
WKB formula under the condition
\begin{eqnarray}
&&\lambda
  ={1\over 2}{m \epsilon^2\over \alphaa\hbar}\left(1+{{\bfp}_i^2\over m^2\epsilon^2}\right)\gg1,
\label{wkbapp}
\\
&&{p_i\over m\epsilon}\hat k \gg {1\over 2}{\hbar \alphaa\over m\epsilon^2}\hat k^2.
\label{class}
\end{eqnarray}
We can show that the former condition (\ref{wkbapp}) is derived from the condition 
for the WKB approximation, Eq.(\ref{wkbcondition}), while 
the latter condition (\ref{class}) is equivalent to $p_i\gg \hbar k$, i.e.,
Eq.(\ref{classicalpic}).
The former condition is satisfied when the Compton wavelength of 
the charged particle is shorter than the Hubble horizon length 
around the bounce regime defined by $a\sim\epsilon\sim\alphaa\eta$ 
(see below), where the classical radiation rate becomes maximum. 
Figure~\ref{fig3} plots $F$ as a function of $p_i/m\epsilon$ with fixed as
$\hbar\alphaa/m\epsilon^2=1$, 0.1 and 0.01.
Figure~\ref{fig4} plots $F$ as a function of $\hbar\alphaa/m\epsilon^2$
with fixed as $p_i/m\epsilon=0.01$ and $1$. 
These figures show $F=1$ for $p_i/m\epsilon\simlt 1$ and 
$\hbar\alphaa/m\epsilon^2\ll 1$, and the 
suppression $F<1$ for the other region. 

Finally in this section, we mention the physical meaning of these 
parameters. 
Around the bounce regime $a\sim\epsilon\sim\alphaa\eta$, we can write
\begin{eqnarray}
  {p_i\over m\epsilon}\simeq{p_{phys}\over m}\Big|_{\rm bounce}, \qquad
    {\hbar\alphaa\over m\epsilon^2}\simeq\lambda_{\rm C}{H}\Big|_{\rm bounce},
\end{eqnarray}
where $p_{phys}=p_i/\epsilon$ is the physical momentum, 
$\lambda_{\rm C}=\hbar/m$ is the Compton wavelength, and 
$H|_{\rm bounce}=\alphaa/\epsilon^2$ is the Hubble expansion rate,
 respectively. 
Thus $p_i/m\epsilon$ is the relativistic factor, and 
$\hbar\alphaa/m\epsilon^2$ can be regarded as the
the ratio of the Hubble horizon length to the Compton 
wavelength of the charged particle, around the bouncing regime. 

\subsection{Example II}
\def\bart{{\bar t}}
\def\barm{{\bar m}}
\def\barp{{\bar q}}
\def\bark{{\bar k}}
\def\bareta{{\bar \eta}}

Here we consider a Friedmann spacetime with the scale factor 
\begin{eqnarray}
  a=\left\{
\begin{array}{cc}
  +\bar t+\epsilon & (\bart>0)\\
  -\bar t+\epsilon & (\bart<0)
\end{array}
\right.
  =\left\{
\begin{array}{cc}
  \epsilon  e^{+\bareta}& (\bareta>0)\\
  \epsilon  e^{-\bareta}& (\bareta<0)
\end{array}
\right.,
\label{continue}
\end{eqnarray}
where $\epsilon(>0)$ is a small constant, $\bart =t/t_0$ 
and $\bareta=\eta/\eta_0$. This mimics a Milne-like bounce
universe but with a flat spatial geometry.

The solution of the Klein-Gordon equation is written with the Bessel function, 
and the positive frequency mode function is \cite{NariaiAzuma}
\begin{eqnarray}
  \varphi_{q}=N_\barp\times\left\{
\begin{array}{lc}
  H_{-i\barp}^{(2)}(\barm (+\bart+\epsilon)) \hspace{1cm} & (t>0)\\
  H_{i\barp}^{(1)}(\barm (-\bart+\epsilon))  \hspace{1cm} & (t<0)
\end{array}
\right.,
\end{eqnarray}
where
\begin{eqnarray}
  N_\barp= e^{-\barp\pi/2}{\sqrt{\eta_0\pi}\over 2}
\end{eqnarray}
from the normalization condition, and we have
defined $\barm=\eta_0m/\hbar$ and $\barp=\eta_0 q$. 

The in- and out-vacuum mode functions are therefore given by
\begin{eqnarray}
 \varphi^{\rm in}_{q}
&=&N_\barp\times\left\{
\begin{array}{lc}
  \alpha_\barp H_{-i\barp}^{(2)}(\barm(\bart+\epsilon))
  +\beta_\barp H_{i\barp}^{(1)} (\barm(\bart+\epsilon)) & \hspace{1cm} (\bart>0)\\
  H_{i\barp}^{(1)}(\barm(-\bart+\epsilon))) & \hspace{1cm} (\bart<0)
\end{array}
\right.
\\
\nonumber\\
\varphi^{\rm out}_{q}
&=&N_\barp\times\left\{
\begin{array}{lc}
   H_{-i\barp}^{(2)}(\barm(\bart+\epsilon)))&\hspace{1cm} (\bart>0)\\
   \bar\alpha_\barp H_{i\barp}^{(1)}(\barm(-\bart+\epsilon))
   +\bar\beta_\barp H_{-i\barp}^{(2)} (\barm(-\bart+\epsilon))&\hspace{1cm} 
  (\bart<0)
\end{array}
\right.,
\nonumber\\
\end{eqnarray}
where the coefficients are
\begin{eqnarray}
&&\alpha_\barp={\pi \barm \epsilon \over 4i} e^{-\pi \barp}
(H_{i\barp}^{(1)}(\barm\epsilon))^2{'},
\nonumber\\
&& \beta_\barp=-{\pi \barm \epsilon \over 4i} e^{-\pi \barp}
      (H_{i\barp}^{(1)}(\barm\epsilon)H_{-i\barp}^{(2)}(\barm\epsilon))',
\nonumber\\
&& \bar\alpha_\barp=-{\pi \barm \epsilon \over 4i} e^{-\pi \barp}
                (H_{-i\barp}^{(2)}(\barm\epsilon)){}^2{}'=\alpha_\barp^*,
\nonumber\\
&& \bar\beta_\barp={\pi \barm \epsilon \over 4i} e^{-\pi \barp}
      (H_{i\barp}^{(1)}(\barm\epsilon)H_{-i\barp}^{(2)}(\barm\epsilon))'
=\beta_{\barp}^*,
\end{eqnarray}
and the prime means $H(z)'=dH(z)/dz$. 

We may write
\begin{eqnarray}
  e^{i\bark\bareta}= \left\{
\begin{array}{lc}
   (\barm\epsilon)^{-i\bark} (\barm(\bart+\epsilon))^{ i\bark} 
  & \hspace{1cm} (\bart>0),\\
   (\barm\epsilon)^{i\bark}  (\barm(-\bart+\epsilon))^{-i\bark} 
  & \hspace{1cm} (\bart<0),
\end{array}
\right.
\end{eqnarray}
then, we have
\begin{eqnarray}
  &&\int_{-\infty}^\infty d\eta e^{ik\eta} \varphi_{q_f}^{\rm in*} 
  \varphi_{q_i}^{\rm out}
\nonumber\\
  &&= \eta_0 N_{\barp_f}N_{\barp_i}(\barm \epsilon)^{-i\bark}
  \int_{\barm\epsilon}^\infty dz z^{i\bark -1} 
  \left(\alpha_{\barp_f}^*H_{i\barp_f}^{(1)}(z)
              +\beta_{\barp_f}^*H_{-i\barp_f}^{(2)}(z)\right)
  H_{-i\barp_i}^{(2)}(z) 
\nonumber \\
  &&+\eta_0 N_{\barp_f}N_{\barp_i}(\barm \epsilon)^{i\bark}
  \int_{\barm\epsilon}^\infty dz z^{-i\bark -1} 
  H_{-i\barp_f}^{(2)}(z) 
  \left(\bar\alpha_{\barp_i}H_{i\barp_i}^{(1)}(z)
              +\bar\beta_{\barp_i}H_{-i\barp_i}^{(2)}(z)\right).
\nonumber
\\
\end{eqnarray}

In the limit $z\gg1$, we have
\begin{eqnarray}
&& H_{ip}^{(1)}(z)=\sqrt{2\over\pi z} e^{iz+p\pi/2-i\pi/4},
\nonumber\\
&& H_{-ip}^{(2)}(z)=\sqrt{2\over\pi z} e^{-iz+p\pi/2+i\pi/4}.
\nonumber
\end{eqnarray}
Thus in the limit, $\barm \epsilon=m\eta_0\epsilon/\hbar \gg1$, 
we may approximate 
\begin{eqnarray}
&&\alpha_{\barp_i}=\bar\alpha_{\barp_f}^*\simeq-i e^{2i\barm\epsilon},
\\
&&\beta_{\barp_i}=\bar\beta_{\barp_f}^*\simeq0.
\end{eqnarray}
and we have
\begin{eqnarray}
  \int_{-\infty}^\infty d\eta e^{ik\eta} \varphi_{q_f}^{\rm in*} 
  \varphi_{q_i}^{\rm out}
  &=& -i e^{2i\barm\epsilon}{\eta_0^2\over \barm \epsilon (1+\bark^2)}.
\end{eqnarray}
After the integration with respect to $\cos\theta$, the total energy is 
\begin{eqnarray}
 {E}={2e^2 q_i^2 \eta_0 \over 3\pi^2\varepsilon_0 \barm^2 \epsilon^2 }
 \int_0^\infty d\bark {\bark^2\over (1+\bark^2)^2}={e^2 p_i^2 \over 6\pi\varepsilon_0 \eta_0 m^2 \epsilon^2 }.
\label{modeEE}
\end{eqnarray}
This completely agrees with what is obtained 
from the integration of (\ref{feneb}) with (\ref{continue}).
Note that this result is obtained under the condition
$\barm \epsilon=m\eta_0\epsilon/\hbar \gg1$, which may be expressed as 
\begin{eqnarray}
 \barm \epsilon=m\eta_0\epsilon/\hbar = \lambda_{\rm C}^{-1}{H^{-1}}\big|_{\rm bounce}\gg1,
\end{eqnarray}
where $\lambda_{\rm C}=\hbar/m$ is the Compton wavelength and 
$H|_{\rm bounce}=1/(\epsilon\eta_0)$ is the Hubble expansion 
rate at the bounce regime $a\sim\epsilon$, i.e, $(|\eta|\simlt\eta_0)$.
Thus the WKB formula is valid as long as the Compton wavelength is 
shorter than the Hubble horizon length. 

\section{Summary and Conclusions}
In the present paper, we investigated photon emission from a
moving massive charge in an expanding universe.
We considered the scalar QED model for simplicity, and
focused on the energy radiated by the process.
First we showed how the Larmor formula for the rate of the 
radiation of energy in the classical electromagnetic theory
can be reproduced under the WKB approximation
in the framework of the quantum field theory in curved 
spacetime.

We also investigated the limits of the validity of the WKB
formula, by deriving the radiation formula in a bouncing universe
in which the mode functions
are exactly solvable. The result using the exact mode function
shows the suppression of the radiation energy compared with the WKB formula.
The suppression depends on the ratio of 
the Compton wavelength $\lambda_C$ of the charged particle to 
Hubble length $H^{-1}$.
Namely, the larger the ratio $\lambda_C/H^{-1}$ is,
the stronger the suppression becomes.
In the limit the Compton wavelength is small compared with
the Hubble length, the radiation formula is found to agree with
the WKB formula. Since this limit is equivalent to
the limit $\hbar\rightarrow0$, the suppression we found 
is a genuine quantum effect in an expanding (or contracting) 
universe, which is due to the finiteness of the Hubble length.
Whether the quantum effect on the radiation from a accelerated 
charge always leads suppression or not is an interesting question.
This would be understood by analyzing higher order terms of 
the WKB approximation. We will return to this point in future.

\vspace{2mm}
{\it Acknowledgments}~
This work was supported by Grant-in-Aid of Science research of Japanese
Ministry of Education, Culture, Sports, Science and Technology (Nos. 18654047, 18540277, 14102004, 17340075, and 18204024). 
All the numerical computation presented in this paper were 
performed with the help of the package MATHEMATICA version 
5.1. We thank Y.~Kojima, T.~Takahashi, K.~Homma, K.~Yokoya, H.~Okamoto, 
A.~Higuchi, H.~Sato, J.~Yokoyama, T.~Morozumi, and M.~Okawa 
for useful communications related to the topic of the present 
work. 

\section*{References}

\end{document}